\documentclass[a4paper]{jpconf}
\usepackage{graphicx}
\usepackage{amsmath,amssymb}
\usepackage{soul,color}
\usepackage[dvipsnames]{xcolor}
\usepackage{slashed}
\begin{document}
\title{Casimir energy on the sphere and 6D CFT trace anomaly}

\author{R. Aros$^1$, F. Bugini$^2$, D.E. Díaz$^3$ and C. N\'u\~nez-Barra$^4$}

\address{$^1$ Departamento de Ciencias Fisicas, Universidad Andres Bello,
Sazie 2212, Piso 7, Santiago, Chile}
\address{$^2$ Departamento de Matem\'atica y F\'isica Aplicadas, 
Universidad Cat\'olica de la Sant\'isima Concepci\'on,
Alonso de Ribera 2850, Concepci\'on, Chile}
\address{$^3$ Departamento de Ciencias Fisicas, Universidad Andres Bello,
Autopista Concepcion-Talcahuano 7100, Talcahuano, Chile}
\address{$^4$ Facultad de Física, Pontificia Universidad Católica de Chile, Vicuña Mackenna 4860, Santiago, Chile}
\ead{raros@unab.cl,fbugini@ucsc.cl,danilodiaz@unab.cl,cnb@uc.cl}
\begin{abstract}
 We elucidate the dependence of the Casimir energy on the trace anomaly coefficients for a six-dimensional CFT on $R\times S^5$. This extends the universal dependence on the central charge in 2D and the relation by Cappelli and Coste in 4D, unveiling the role of the trivial total derivatives in the anomaly that render the Casimir energy scheme dependent. We obtain 
 $$E_o=-\frac{15}{8}\,a_6 -\frac{5}{12}\,\left(g_5+\frac{1}{4}\,g_7+\frac{1}{2}\,g_8-10\, g_9+g_{10}\right),$$
 with $a_6$ being the type A central charge and the $g$'s, the coefficients of five out of six terms that form a basis for trivial total derivatives. The derivation is based on the Polyakov formulas (conformal primitive) resulting from the integration of the trace anomaly. \\ 
 Alternatively, on a 6D conformally flat background the above basis is redundant and one can simplify further to get, in terms of the Schouten scalar $J$ and the Schouten tensor $V$, Branson's basis  for trivial total derivatives  $\nabla^2\nabla^2 J$, $\nabla^2J^2$ and $\nabla^2|V|^2+2\nabla\cdot(V\cdot\nabla\,J)$ with coefficients $\gamma_1, \gamma_2$ and $\gamma_3$, respectively,    
\begin{equation}
\nonumber
    E_o=-\frac{15}{8}a_6-\frac{1}{24}\left(\gamma_1-\gamma_2 -\frac{1}{8}\gamma_3\right)~.
\end{equation} 
\end{abstract}

\section{Introduction}
The Casimir effect~\cite{Casimir48} is a remarkable macroscopic quantum response of a system to external influences (e.g. boundary conditions or background fields). It has been extensively studied over the years and experimentally verified~\cite{Sparnaay58, Lamoreaux98} (for a recent review, see \cite{Dantchev:2022hvy}). One interesting route contemplates the extension of the original setting regarding the 4D electromagnetic field to a conformal field theory in a generic even dimension. 
Since conformal symmetry is quite restrictive, the Casimir energy of a 2D CFT on the cylinder is universal and given by the {\em ubiquitous} central charge $c$~\cite{Bloete:1986qm, Affleck:1986bv} of the trace anomaly,
\begin{equation}
    E_o=-\frac{c}{12}.
\end{equation}
The aforementioned universality is partially lost in 4D,  yet the trace anomaly coefficients determine the vacuum energy\footnote{We stick to the conventions of~\cite{Beccaria:2014xda}.} as shown by Cappelli and Coste~\cite{CAPPELLI1989707}      
\begin{equation}
    E_o=\frac{3}{4}\,a+\frac{3}{8}\,g~,
\end{equation}
where $a$ is the universal type A trace anomaly coefficient and $g$ is the regularization scheme dependent coefficient of the total derivative term 
\begin{equation}
    (4\pi)^2 \langle T\rangle = -a\,E_4 + c\,W^2 + g\,\nabla^2 R~.
\end{equation}
The total derivative term $\nabla^2 R$ is a {\em trivial} anomaly since it comes from the conformal variation~\footnote{By conformal we mean a local Weyl rescaling of the metric. We will use both terms, conformal and Weyl, indistinctly in what follows.} of a conformal primitive that happens to be a local curvature invariant with the appropriate scaling dimension\footnote{The other two candidates for a local curvature invariant $Riem^2$ and $Ric^2$ can be reduced to $R^2$ by exploiting the conformal invariance of the Euler density $E_4$ and of the quadratic contraction of the Weyl tensor $W^2$.}, namely $R^2$, that can simply be added to the action.

There are good indications that in higher (necessarily even) dimensions the Casimir energy on the cylinder~\footnote{Meaning that the spatial section is compactified to a sphere.} is still dictated by the trace anomaly of the CFT. The structure of the latter was asserted years ago by Deser and Schwimmer~\cite{Deser:1993yx} to consist of~\footnote{See also~\cite{Bonora1983,Boulanger:2007st,Boulanger:2007ab} for a cohomological approach. For the closely related question of the classification of global conformal invariants, see~\cite{Branson1995} on conformally flat manifolds and~\cite{Alexakis2012,Boulanger:2018rxo} in greatest generality.} a multiple of the Euler density (type A), a point-wise Weyl invariant (type B) and a {\em trivial} total derivative       
\begin{eqnarray}
 \nonumber (4\pi)^{\frac{n}{2}}\cdot\langle T \rangle &=& - (-)^{\frac{n}{2}}\, a_n \cdot E_n + \left\{\mbox{Point-wise Weyl Invariant}\right\} + \mbox{t.t.d.}
\end{eqnarray}
Remarkably, a simple relation between the Casimir energy on the cylinder and the type A trace anomaly coefficient was found by Herzog and Huang in~\cite{Herzog:2013ed}
\begin{equation}
E_o=\frac{(-1)^{\frac{n}{2}}\cdot (n-1)!!}{2^{\frac{n}{2}}}\cdot a_n~,
\end{equation}
but only valid in a particular regularization scheme where the trivial total derivatives are absent. For example, in 6D one encounters discrepancies with the Casimir energy for the free (2,0) tensor multiplet as computed via standard $\zeta$-function regularization (see, e.g.~\cite{Gibbons:2006ij,Beccaria:2014qea}). A 6D relation, like the Cappelli and Coste one, taking into account the role of the trivial total derivatives in the trace anomaly, is still missing. This note aims to provide such an extension. 

Our results are primarily based on a rather simple observation: since the trivial total derivatives in the trace anomaly stem from a local conformal invariant in the action, it is enough to evaluate the latter on the geometry of the cylinder to obtain its contribution to the partition function and, in consequence, to the vacuum energy. But this observation alone does not produce the precise relation between vacuum energy and trace anomaly that we are looking for, it merely asserts that both quantities are subject to ambiguities depending on the renormalization scheme. However, and this is the key to our derivation, when applied to the Polyakov formula  
that provides the conformal primitive of the trace anomaly it brings together the type A central charge $a$ and the trivial total derivative terms resulting in the appropriate extension of the Cappelli-Coste relation to higher dimensions. It is the distinguished feature of Branson's construction, where the Euler density is traded in for the Q-curvature, that reduces the answer to terms at most quadratic in the Weyl factor relating the standard metric on the cylinder to that of a flat shell. Having control on the trivial total derivative terms, one can then go back to the Euler basis and complete the partial result by Herzog and Huang provided a dictionary between critical Q-curvature and Euler density is at hand. This dictionary, in particular on the conformally flat class of background metrics, has been established by Graham and Juhl~\cite{Graham:2007} in the Conformal Geometry literature after the advent of the AdS/CFT correspondence in Physics. 

Although our derivation is conceptually clear, in practice we face the complication that, unlike the Euler term, the number of independent trivial total derivatives increases significantly with the dimension; there are none in 2D, one in 4D, six in 6D, whereas 8D and beyond remain largely unexplored~\footnote{There is, though, a basis of nine local conformal primitives on conformally flat backgrounds in 8D due to Branson and Peterson~\cite{Branson2007}.}.
We start in Section 2 with a brief account of Polyakov formulas. We then consider the explicit cases of 2D, 4D, and 6D in Section 3 and test in Section 4 against data available in the literature. We then conclude and collect some useful relations in the appendix.  

\section{Polyakov formulas, Q-curvature, and local curvature invariants}

The (generalized) Polyakov formula provides a conformal primitive of the trace anomaly, as first shown by Polyakov in 2D~\cite{Polyakov:1981rd}. In higher even dimensions, Branson asserted the form of the corresponding extension by making use of the suitable features of the critical Q-curvature (see, e.g.~\cite{BransonSharp1995}). In a (closed Riemannian) conformally flat manifold the topological content of the trace anomaly is also captured by the Q-curvature and all point-wise Weyl invariants vanish so that the structure of the trace anomaly is significantly simplified            
\begin{eqnarray}
  (4\pi)^{\frac{n}{2}}\cdot\langle T \rangle &=& - (-)^{\frac{n}{2}}\, n\cdot a_n \cdot Q_n + \mbox{t.t.d.}
\end{eqnarray}
Here one exploits the linear dependence of the variation of the Q-curvature on the Weyl factor $\sigma$, that conformally relates two metrics $\hat{g}=e^{2\,\sigma}g$,  and integrates in an auxiliary parameter $t$ along any interpolating path $\sigma(t)$ such that $\sigma(0)=0$ and $\sigma(1)=\sigma$, to obtain for the partition functions
\begin{eqnarray}
\label{Dilatonaction}
    \nonumber\Gamma[g,\hat{g}]=-\Gamma[\hat{g},g]&=&\log Z[\hat{g}]-\log Z[g] =-(-)^{\frac{n}{2}}\, \frac{n\cdot a_n}{(4\pi)^{\frac{n}{2}}} \int_{M_n} dv_g \;\sigma\cdot\left\{ Q_n + \frac{1}{2}\,P_n\,\sigma\right\} \\\nonumber\\
        &&
    +\left\{\int_{\hat{M}_n} dv_{\hat{g}} \;\hat{\mathcal{F}}-\int_{M_n} dv_{g} \;{\mathcal{F}} \right\}~.
 \end{eqnarray}
The explicit dependence on $\sigma$ is at most quadratic and the GJMS operator $P_n$ can be integrated by parts to yield the so-called dilaton action for the Weyl factor, whereas higher order terms in $\sigma$ are hidden in the finite conformal variation of the local conformal primitive $\mathcal{F}$.  

Now, considering the standard metric on the cylinder $R\times S^{n-1}$ conformally related to flat space 
\begin{equation}
    \hat{g}=du^2+d\Omega^2=r^{-2}\left\{dr^2+r^2\,d\Omega^2\right\}
\end{equation}
via a Weyl factor $\sigma=-\ln r$, we can readily read off the Casimir energy resulting from the compactification of the spatial section referred to as the vanishing vacuum energy of flat space. Alternatively, one can consider a temperature circle with inverse temperature $\beta$ and read off the dominant vacuum term as $\beta$ goes to infinity on $S^1_{\beta}\times S^{n-1}$.\\  
In ``flatland'' (meaning flat space or a spherical shell for finite inverse temperature), the only contribution comes from the nth-order kinetic term $|\partial\partial^2...\partial^2\sigma|^2$ or $|\partial^2...\partial^2\sigma|^2$ depending on the (even) dimension $n$, that produces the same answer $((n-2)!!)^2/r^{n}$ so that the radial integral results in $\int \frac{dr}{r}=\int du$ or $\beta$. Performing the remaining volume integral on the sphere, the generic relation for the Casimir energy (modulo trivial total derivative contributions) becomes
\begin{equation}
E_o=\frac{(-1)^{\frac{n}{2}}\cdot n!!}{2^{\frac{n}{2}+1}}\cdot a_n+...~.
\end{equation}
This ought to be compared with the result by Herzog and Huang in the Euler density basis
\begin{equation}
E_o=\frac{(-1)^{\frac{n}{2}}\cdot (n-1)!!}{2^{\frac{n}{2}}}\cdot a_n+...~.
\end{equation}
The apparent discrepancy above stems from the fact that each result is valid within a different regularization scheme where trivial total derivatives are discarded. However, evaluation of the local conformal primitive of the trivial total derivatives on the cylinder will complement the result and the agreement between the two schemes and a more general one will be achieved. The local conformal primitive vanishes on flat space, and must be a constant on the cylinder. This yields the scheme-dependent contribution to the Casimir energy foreseen above. 
The only technical difficulty in carrying out the above completion, as already pointed out, lies in the increasing complexity of the trivial total derivative terms and their conformal primitives with the dimension.

\section{Casimir energy and trace anomaly: the explicit relation}
In what follows we will revisit the 2D and 4D cases and obtain, for the first time to our knowledge, the 6D extension of the Capelli-Coste relation. 
\subsection{2D CFT}
Let us start by reproducing the 2D relation~\footnote{This was essentially obtained by Affleck~\cite{Affleck:1986bv}, except for the fact that his result applies to the modular transformed torus, i.e., to the Cardy or high temperature regime.} following Cappelli and Coste. There are no trivial anomalies in 2D, by convention $Q_2=\frac{R}{2}$ is the Gaussian curvature of the closed surface and $P_2=-\nabla^2$, that is integrated by parts to write down the answer in the well-known Liouville form 
\begin{equation}
    \Gamma[g,\hat{g}]=\frac{a_2}{4\pi} \int_{\text{flatland}} dv_g \;\left\{\; |\nabla\sigma|^2 + \sigma\cdot R  \right\} 
\end{equation} 
From this potential for the Weyl factor, also known in the conformal geometry literature as the Moser functional, evaluated on the fiducial flat metric with $R=0$ and for the conformal factor $|\partial\sigma|^2=\frac{1}{r^2}$, one can readily obtain the leading contribution to the partition function as $\beta\rightarrow\infty$
\begin{equation}
-\beta\cdot E_o=\frac{a_2}{4\pi}\cdot2\pi\cdot\beta=\frac{a_2}{2}\cdot\beta
\end{equation}
This corresponds to the more familiar $E_o=-c/12$, when the central charge is tuned so that for the scalar Laplacian $a_2=1/6$ one gets $c=1$, i.e., $a_2=c/6$. In all, 
\begin{equation}
E_o=-\frac{a_2}{2}.
\end{equation}

\subsection{4D CFT}
Now we revisit the derivation by Cappelli and Coste in 4D, but in the light of Branson's insight that exploited the suitable transformation law of the Q-curvature under Weyl rescaling of the metric (see also Riegert's action~\cite{Riegert:1984kt} in local form). The trace anomaly at a conformally flat background becomes
\begin{equation}
(4\pi)^2\cdot\langle T \rangle = -a_4 \cdot E_4 + g\cdot\nabla^2\,R = -4\,a_4\cdot Q_4 + \tilde{g}\cdot\nabla^2\,R~, 
\end{equation}
and its conformal primitive, referred to the fiducial flat metric, is then given by  
 \begin{eqnarray}
 \label{DilatonAction}
    \nonumber\Gamma[g,\hat{g}]&=&-\frac{a_4}{4\pi^2}\int_{\text{flatland}} dv_g\left\{\sigma\cdot Q_4 + \frac{1}{2}\,\sigma\cdot P_4\,\sigma\right\}+\frac{\tilde{g}}{16\pi^2}\int_{S^1_{\beta}\times S^3} dv_{\hat{g}} \;\frac{-1}{12}\hat{R}^2
    \\\nonumber\\
        &=&-\frac{a_4}{8\pi^2}\int_{\text{flatland}} dv_g\;|\partial^2\sigma|^2+\frac{\tilde{g}}{16\pi^2}\int_{S^1_{\beta}\times S^3} dv_{\hat{g}} \;\frac{-1}{12}\hat{R}^2~.
\end{eqnarray} 
We obtain contribution from the fourth-order kinetic term $|\partial^2\sigma|^2=\frac{4}{r^4}$ and from the local curvature invariant $\frac{-1}{12}\hat{R}^2=-3$, conformal primitive of the trivial total derivative, so that we end up with
\begin{equation}
-\beta\cdot E_o=-\left\{\frac{a_4}{2\pi^2}+\frac{3\tilde{g}}{16\pi^2} \right\}\cdot 2\pi^2\cdot\beta=-\left\{a_4+\frac{3}{8}\,\tilde{g}\right\}\cdot\beta~.
\end{equation}
Therefore, we obtain for the Casimir energy 
\begin{equation}
E_o=a_4+\frac{3}{8}\,\tilde{g}~.
\end{equation}
To translate to the Euler basis, it is enough to remember that on conformally flat backgrounds $4Q_4=E_4-\frac{2}{3}\nabla^2 R$, so that $\tilde{g}=g-\frac{2}{3}a_4$ and finally obtain the original result by Cappelli and Coste\footnote{Our convention follows~\cite{Beccaria:2014xda} and slightly differs from that of the original Cappelli-Coste paper~\cite{CAPPELLI1989707}.}
\begin{equation}
E_o=\frac{3}{4}\, \left(a_4+\frac{1}{2}\, g\right)~.
\end{equation}
Notice that the precise shift $\tilde{g}=g-\frac{2}{3}a_4$ decorates the dilaton action (eq.~\ref{DilatonAction}), by using the conformal transformation of $\sqrt{\hat{g}}\hat{R}^2$, as follows
\begin{eqnarray}
\nonumber\Gamma[\delta,\hat{\delta}]&=&-\frac{a_4}{8\pi^2}\int_{\text{flatland}} dv_{\delta}\;|\partial^2\sigma|^2-\frac{2}{3}\frac{a_4}{16\pi^2}\int_{\text{flatland}} dv_{\delta} \;\frac{-1}{12}\left\{-6\partial^2\sigma-6|\partial\sigma|^2\right\}^2\,+...\\\nonumber\\
&=&\frac{a_4}{8\pi^2}\int_{\text{flatland}} dv_{\delta}\;\left\{2|\partial\sigma|^2\partial^2\sigma + (|\partial\sigma|^2)^2 \right\}\,...~,
\end{eqnarray}
where the ellipsis stands for the contribution of the trivial anomaly $g\nabla^2R$. Evaluating the first part on the cylinder will again produce the dependence of the vacuum energy on the central charge $a$, while the contribution from the trivial anomaly can easily be reintroduced by evaluating the associated counterterm (conformal primitive)~\footnote{ This shortcut route to the Cappelli-Coste relation was somehow already implicit in~\cite{Assel:2015nca}(see appendix A.1 therein, and also~\cite{Benjamin:2023qsc}).}.
\subsection{6D CFT}

It is at 6D that a subtle departure between total derivatives and {\em trivial} total derivatives occurs, stemming out from the difference between global conformal invariants and the trace anomaly of a CFT. The latter must comply with the Wess-Zumino consistency or integrability condition, requiring that the allowed total derivatives must be derived from the conformal variation of a local curvature invariant. In 6D, a basis for total derivatives is provided by the seven divergences $C_1, C_2, C_3, C_4, C_5, C_6$ and $ C_7$ of~\cite{Bastianelli:2000hi}, however once they are expressed in the reduced basis for trivial total derivatives of $M_5, M_6, M_7, M_8, M_9$ and $M_{10}$ there remains an ``offending term" $\nabla^2 |Ric|^2$. The basis for trivial total derivatives was obtained in~\cite{Bastianelli:2000rs} by performing the conformal variation of all possible curvature invariants with the appropriate scaling dimension. There are seventeen elements $K_1, ..., K_{17}$, from which seven combinations are the total derivatives $C_1, C_2, C_3, C_4, C_5, C_6, C_7$ whose integrals vanish identically, another combination gives the Euler density $E_6$ which is a topological invariant, and there are also three other combinations that produce three independent point-wise conformal invariants $I_1, I_2$ and $I_3$, so that in all there remain six independent combinations -conformal primitives that we denote $F_5, F_6, F_7, F_8, F_9, F_{10}$ -   whose conformal variations produce the trivial total derivatives $M_5, M_6, M_7, M_8, M_9, M_{10}$ (we collect explicit expressions in~\ref{appendix A}). 

Therefore, the 6D CFT trace anomaly at a conformally flat background becomes
\begin{eqnarray}
\nonumber (4\pi)^3\cdot\langle T \rangle &=& a_6 \cdot E_6 + g_5\cdot M_5 + g_6\cdot M_6 + g_7\cdot M_7 + g_8\cdot M_8 + g_9\cdot M_9 + g_{10}\cdot M_{10}\\&=&6\,a_6 \cdot Q_6 + \tilde{g}_5\cdot M_5 + \tilde{g}_6\cdot M_6 + \tilde{g}_7\cdot M_7 + \tilde{g}_8\cdot M_8 + \tilde{g}_9\cdot M_9 + \tilde{g}_{10}\cdot M_{10}~.
\end{eqnarray}
The conformal primitive, referred to a fiducial flat metric where all curvature invariants vanish, is then given by  
 \begin{eqnarray}
    \Gamma[g,\hat{g}]&=&\frac{3\,a_6}{64\pi^3}\int_{\text{flatland}} dv_g \;|\partial\,\partial^2\sigma|^2\\\nonumber\\\nonumber
    &&+\frac{1}{64\pi^3}\int_{S^1_{\beta}\times S^5} dv_g\left\{\tilde{g}_5\cdot \mathcal{\hat{F}}_5+{\tilde{g}_6}\cdot \mathcal{\hat{F}}_6+{\tilde{g}_7}\cdot \mathcal{\hat{F}}_7+{\tilde{g}_8}\cdot \mathcal{\hat{F}}_8+{\tilde{g}_9}\cdot \mathcal{\hat{F}}_9+{\tilde{g}_{10}}\cdot \mathcal{\hat{F}}_{10}\right\}.
\end{eqnarray} 
When evaluated at $S^1_{\beta}\times S^5$ with the standard product metric, we obtain contributions from the sixth-order kinetic term $|\nabla\,\nabla^2\sigma|^2=\frac{64}{r^6}$ and from the local curvature invariants 
\begin{eqnarray}
\nonumber
    \mathcal{\hat{F}}_5=\frac{80}{3}\quad,\quad \mathcal{\hat{F}}_6=0\quad,\quad \mathcal{\hat{F}}_7=\frac{20}{3}\quad,\\\nonumber\\
\mathcal{\hat{F}}_8=\frac{40}{3}\quad,\quad \mathcal{\hat{F}}_9=-\frac{800}{3}\quad,\quad \mathcal{\hat{F}}_{10}=\frac{80}{3}\quad.
\end{eqnarray}
Collecting all contributions 
\begin{eqnarray}
\nonumber
-\beta\cdot E_o&=&\frac{1}{64\pi^3}\left\{3a_6\cdot64 +\tilde{g}_5\cdot\frac{80}{3}+\tilde{g}_7\cdot\frac{20}{3}+\tilde{g}_8\cdot\frac{40}{3}+\tilde{g}_9\cdot\frac{-800}{3}+\tilde{g}_{10}\cdot\frac{80}{3} 
  \right\}\cdot \pi^3\cdot\beta\\
&=&\left\{3a_6+\frac{5}{12}\,\left[\tilde{g}_5+\frac{1}{4}\,\tilde{g}_7+\frac{1}{2}\,\tilde{g}_8-10\,\tilde{g}_9+\tilde{g}_{10}\right]\right\}\cdot\beta~.
\end{eqnarray}
Therefore, we obtain for the Casimir energy 
\begin{equation}
\label{Ec6D}
E_o=-3\,a_6-\frac{5}{12}\,\left[\tilde{g}_5+\frac{1}{4}\,\tilde{g}_7+\frac{1}{2}\,\tilde{g}_8-10\,\tilde{g}_9+\tilde{g}_{10}\right]~.
\end{equation}
To translate to the Euler basis we have to express the 6D Q-curvature in terms of the Euler density and trivial total derivatives in a conformally flat background. The linearity under conformal transformations, a defining property of the Q-curvature, is shared by the whole family of {\it pondered Euler densities} $\tilde{E}_6$ introduced by Anselmi~ in \cite{Anselmi:1999uk} (see \cite{Bastianelli:2000rs} as well). In our conventions, the dictionary goes as follows
\begin{equation}
\label{Anselmi}
    E_6=\tilde{E}_6 +\left(\frac{67}{10}-\frac{5}{2}\zeta\right)M_6+\left(\frac{5}{2}\zeta-\frac{51}{5}\right)M_7+\left(\frac{1}{16}\zeta-\frac{9}{200}\right)M_9-\frac{3}{5}M_{10}~,
\end{equation}
and results in the shift of the trivial anomaly coefficients
\begin{eqnarray}
    \tilde{g}_5=g_5\quad&,&\quad\tilde{g}_6=g_6+a_6\left(\frac{67}{10}-\frac{5}{2}\zeta\right)\quad,\quad\tilde{g}_7=g_7+a_6\left(\frac{5}{2}\zeta-\frac{51}{5}\right)\quad,\nonumber\\    \tilde{g}_8=g_8\quad&,&\quad\tilde{g}_9=g_9+a_6\left(\frac{1}{16}\zeta-\frac{9}{200}\right)\quad,\quad\tilde{g}_{10}=g_{10}-\frac{3}{5}a_6~.
\end{eqnarray}
Substituting back in the Casimir energy relation (eqn.\ref{Ec6D}), we obtain the alternative relation that completes the partial result obtained by Herzog and Huang in 6D
\begin{equation}
E_o=-\frac{15}{8}\,a_6-\frac{5}{12}\,\left[{g}_5+\frac{1}{4}\,{g}_7+\frac{1}{2}\,{g}_8-10\,{g}_9+{g}_{10}\right]~.
\end{equation}
The above relation is robust in the sense that any construct with the same conformal transformation property as the critical Q-curvature $Q_6$ will produce the same result.
For completeness, let us also examine the two-parameter family of trivial total derivative terms that respects the linearity in a generic background~\footnote{We are grateful to I.L. Shapiro for a useful discussion on this point.} found out by Hamada~\cite{Hamada:2000me} (eqn.4.20 and 4.21 therein, see also~\cite{Karakhanian:1994yd,Ferreira:2018utt}) at the expense of modifying Branson's operator $P_6$~:
\begin{eqnarray}
   \label{Hamada1} &(i)&\frac{3}{4}\nabla^2W^2-3\nabla\nabla(WW)\,=\,-\frac{3}{4}C_5+3C_7\,=\,\frac{1}{4}M_5+\frac{1}{2}M_6-\frac{3}{4}M_7+M_8+\frac{9}{160}M_9\\
    \nonumber\\
    \label{Hamada2}&(ii)&\frac{3}{2}\nabla^2W^2\,=\,3C_5\,=\,-3M_6+3M_8+\frac{3}{20}M_9
\end{eqnarray} 
It is reassuring to verify that none of the two combinations, $\nabla\nabla(WW)$ nor $\nabla^2 W^2$, will produce a non-vanishing result in the relation for the Casimir energy. The above freedom agrees with the usual wisdom in Conformal Geometry that the pair of critical Q-curvature $Q_n$ and critical GJMS operator $P_n$ is ``pure Ricci'' , while any other construct that shares the same conformal properties is ``contaminated'' by explicit dependence on the Weyl tensor.

\section{6D CFT: examples}
There are few examples in 6D, worked out some years ago~\cite{Bastianelli:2000hi}, that can now be used to verify the correctness of the novel relation put forward above between Casimir energy and trace anomaly coefficients. We first write down the trace anomaly with the relevant coefficients, and in the table below we report the values for the Casimir energies computed with the relation we found out and the outcomes in all cases exactly match the known results in the literature (see, e.g., \cite{Gibbons:2006ij}).

\paragraph{Conformal scalar:}
The first instance is of course the conformally coupled scalar field, for which we have
\begin{equation}
     (4\pi)^3\cdot7!\,\langle T \rangle = \frac{5}{9}\cdot E_6 -\frac{28}{3}\cdot I_1+\frac{5}{3}\cdot I_2+2\cdot I_3+\nabla_i\cdot J^i~,
\end{equation}
with the trivial total derivative term given by 
\begin{eqnarray}
     \nabla_i\cdot J^i &=& \frac{6}{5}\cdot C_1 -\frac{2}{5}\cdot C_2+4\cdot C_3+\frac{12}{5}\cdot C_4+\frac{17}{5}\cdot C_6+12\cdot C_7~,\nonumber\\\nonumber\\
     &=& M_5 - 4\cdot M_7 + 7\cdot M_8 +\frac{1}{10}\cdot M_9 + \frac{6}{5}\cdot M_{10}~.
\end{eqnarray}

\paragraph{Massless Dirac squared:}
The second instance of a conformally invariant free field corresponds to the (square of the) massless Dirac operator, for which we have
\begin{equation}
     (4\pi)^3\cdot7!\,\langle T \rangle = \frac{191}{9}\cdot E_6 -\frac{896}{3}\cdot I_1-32\cdot I_2+40\cdot I_3+\nabla_i\cdot J^i~,
\end{equation}
with the trivial total derivative term given this time by 
\begin{eqnarray}
     \nabla_i\cdot J^i &=& 24\cdot C_1 -\frac{148}{15}\cdot C_2+136\cdot C_3+48\cdot C_4-168\cdot C_5 +96\cdot C_6+352\cdot C_7~,\nonumber\\\nonumber\\
     &=&\frac{88}{3}\cdot M_5 - 136\cdot M_7 + \frac{112}{3}\cdot M_8 -5\cdot M_9 + 24\cdot M_{10}~.
\end{eqnarray}

\paragraph{Two-form:}
The third instance of a conformally invariant free field corresponds to a two-form after a covariant gauge-fixing, for which we have
\begin{equation}
     (4\pi)^3\cdot7!\,\langle T \rangle = 442\cdot E_6 -\frac{8008}{3}\cdot I_1-\frac{2378}{3}\cdot I_2+180\cdot I_3+\nabla_i\cdot J^i~,
\end{equation}
and with the trivial total derivative term  
\begin{eqnarray}
     \nabla_i\cdot J^i &=& -60\cdot C_1 +\frac{2036}{15}\cdot C_2-1152\cdot C_3-120\cdot C_4-504\cdot C_5 - 646\cdot C_6+856\cdot C_7~,\nonumber\\\nonumber\\
     &=&\frac{214}{3}\cdot M_5 + 312\cdot M_7 - \frac{14}{3}\cdot M_8 +93\cdot M_9 -60\cdot M_{10}~.
\end{eqnarray}

In summary,

\begin{tabular}{|c|c|c|c|c|c|c|c|c|}
\hline
    & & & & & & & &  \\
    conformal field  & $7!\cdot a$ & $7!\cdot g_5$ & $7!\cdot g_6$ & $7!\cdot g_7$ & $7!\cdot g_8$ & $7!\cdot g_9$ & $7!\cdot g_{10}$ & $E_o$ \\& & & & & & & &  \\\hline
    & & & & & & & & \\  
   conformal scalar  & $\frac{5}{9}$ & 1 & 0 & -4 &  7 & $\frac{1}{10}$ & $\frac{6}{5}$ &  $-\frac{31}{60480}$\\& & & & & & & &  \\\hline
    & & & & & & & &  \\
     massless Dirac squared & $\frac{191}{9}$ & $\frac{88}{3}$ & 0 & -136 &  $\frac{112}{3}$ & -5 & 24 &  $-\frac{367}{24192}$ \\& & & & & & & &  \\\hline
    & & & & & & & &  \\   two-form & 442 & $\frac{214}{3}$ & 0 & 312 & $-\frac{14}{3}$ & 93 & -60 &  $-\frac{191}{2016}$ \\& & & & & & & &  \\\hline
\end{tabular}

\vspace{5mm}
As a verification of the above result for the conformal scalar in 6D, we can evaluate directly in the Polyakov formula obtained by Branson~\cite{BransonSharp1995} (see also \cite{Diaz:2008hy}) 
\begin{eqnarray}
\nonumber
    -&\frac{3\cdot 7!\,(4\pi)^3}{2}&\log\frac{\det\hat{Y}}{\det Y}=10\int_{M_6} dv_g \;\sigma\cdot\left\{ Q_6 + \frac{1}{2}\,P_6\,\sigma\right\} + 13\left\{\int_{\hat{M}_6} dv_{\hat{g}} \;\;|\hat{\nabla}\hat{J}|^2-\int_{M_6} dv_{g} \;|{\nabla}{J}|^2\right\} \\\nonumber\\
        &&+34\left\{\int_{\hat{M}_6} dv_{\hat{g}} \;\hat{J}^3-\int_{M_6} dv_{g} \;J^3\right\}-32\left\{\int_{\hat{M}_6} dv_{\hat{g}} \;\hat{J}\,|\hat{V}|^2-\int_{M_6} dv_{g} \;J\,|V|^2\right\}~,
\end{eqnarray}
in terms of the Schouten tensor $V$ and scalar $J$. Likewise, for the conformally related metrics we get
\begin{eqnarray}
\nonumber
    -{3\cdot 7!\,(4\pi)^3}\,\beta\cdot E_o&=&5\int_{\text{flatland}} dv_g \;|\nabla\,\nabla^2\,\sigma|^2 +34\int_{S^1_{\beta}\times S^5} dv_{\hat{g}}\;\hat{J}^3\,-\,32\int_{S^1_{\beta}\times S^5} dv_{\hat{g}}\;\hat{J}\,|\hat{V}|^2\\\nonumber\\
        &=&\beta\,\pi^3\left\{5\cdot64\,+\,34\cdot 8\,-\,32\cdot3  \right\}\,=\,496\,\beta\,\pi^3~,
\end{eqnarray}
and we again obtain $E_o=-\frac{31}{60480}$.

For the Dirac operator, we start with
\begin{eqnarray}
\nonumber
    -\frac{9\cdot 7!\,(4\pi)^3}{2}\,\log\frac{\det\hat{\slashed{\nabla}}^2}{\det {\slashed{\nabla}}^2}&=&-1146\int_{M_6} dv_g \;\sigma\cdot\left\{ Q_6 + \frac{1}{2}\,P_6\,\sigma\right\}\\ &-& 507\left\{\int_{\hat{M}_6} dv_{\hat{g}} \;\;|\hat{\nabla}\hat{J}|^2-\int_{M_6} dv_{g} \;|{\nabla}{J}|^2\right\} \nonumber\\
        &-& 1578\left\{\int_{\hat{M}_6} dv_{\hat{g}} \;\hat{J}^3-\int_{M_6} dv_{g} \;J^3\right\} \\
        &+&1752\left\{\int_{\hat{M}_6} dv_{\hat{g}} \;\hat{J}\,|\hat{V}|^2-\int_{M_6} dv_{g} \;J\,|V|^2\right\}~ \nonumber, 
\end{eqnarray}
and thus for the conformally related metrics we obtain 
\begin{eqnarray}
\nonumber
    {9\cdot 7!\,(4\pi)^3}\,\beta\cdot E_o&=&-573\int_{\text{flatland}} dv_g \;|\nabla\,\nabla^2\,\sigma|^2 -1578\int_{S^1_{\beta}\times S^5} dv_{\hat{g}}\;\hat{J}^3\,+\,1752\int_{S^1_{\beta}\times S^5} dv_{\hat{g}}\;\hat{J}\,|\hat{V}|^2\\\nonumber\\
        &=&\beta\,\pi^3\left\{-573\cdot64\,-\,1578\cdot 8\,+\,1752\cdot3  \right\}\,=\,-44040\,\beta\,\pi^3~,
\end{eqnarray}
implying $E_o=-\frac{367}{24192}$.

\section{Summary and outlook}
In this work, the relation between the 6D CFT trace anomaly and the Casimir energy due to compactifying the spatial section on a five-sphere has been fully established. The resulting expression extends the 4D result of Cappelli and Coste to 6D by taking account of the trivial total derivatives in the trace anomaly. The scheme-dependence of the  Casimir energy is now precisely controlled by the trivial total derivatives. As it turned out, the partial result by Herzog and Huang obtained via a continuation in dimension argument on the stress-energy tensor is now completed by precisely the evaluation of finite counterterms on the cylinder. There was a priori no guarantee that this would be the final result, for there are ambiguities in the dimensional regularization scheme. For example,  in 4D Brown and Cassidy~\cite{Brown:1977sj} included a vanishing $(n-4)R^2$ that would certainly modify the vacuum energy; in 6D, moreover, the ambiguities grow in complexity due to the many more local curvature invariants that become available. But now, once the trace anomaly is computed in a given renormalization scheme, our completion allows a thorough comparison with the corresponding vacuum energy.

We expect that our 6D version of the Cappelli-Coste relation will allow a better understanding of the discrepancies in the vacuum energy in holographic scenarios. The present relation would also be well suited to further explore the connection, within $\zeta$-function regularization, between Casimir energy and multiplicative anomaly \cite{Aros:2023hgi} beyond conformal scalar fields~\footnote{There is a misprint in eqn. 8.3 of~\cite{Aros:2023hgi}, the coefficient of $\gamma_3$ should be 1 and not 11. The total derivative $\nabla^2|V|^2$ must also be completed to $\nabla^2|V|^2+2\nabla\cdot(V\cdot\nabla J)$ to make it a trivial total derivative.}. \\
Let us conclude by mentioning that going beyond the conformally flat case seems to be quite a challenge (see e.g.~\cite{Huang:2013lhw}). One of the obvious difficulties consists in the choice of a fiducial metric. Even in the conformally flat case, finite size effects on a higher dimensional torus ($d>2$) are not obtained by conformal symmetry from $R^d$ because these geometries are no longer conformally related, as pointed out by Capelli and Coste. Nonetheless, back to the cylinder, a deformation of the sphere could be considered where a suitable candidate for reference metric would be the homogeneously squashed (or Berger), instead of the round, sphere where many explicit calculations are doable (see e.g.~\cite{Zoubos:2004qm}). 
\ack
We thank F. Bastianelli, L. Casarin, J.S. Dowker, E. Friedman, L. Peterson and I.L. Shapiro for valuable conversations and comments. This work was partially funded through FONDECYT-Chile 1220335. We have benefited from the Cadabra software~\cite{Cadabra} to evaluate curvature invariants and the {\em Mathematica} package {\em xAct}~\cite{xAct} to verify their conformal transformations.
\appendix

\section{6D basis of curvature invariants}
\label{appendix A}
A suitable basis for curvature invariants of degree six is due to Fulling et al.~\cite{Fulling:1992vm} 
\begin{eqnarray}
\nonumber
   && K_1=R^3\quad,\quad K_2=R\, Ric^2\quad,\quad K_3=R\, Riem^2\quad,\quad K_4=Ric^3\quad,\quad K_5=-Riem\,Ric^2\\\nonumber \\\nonumber
&&K_6=Ric\,Riem\,Riem\quad,\quad K_7=Riem^3\quad,\quad K_8=Riem'^3\quad,\quad K_9=R\,\nabla^2\,R
\\\nonumber \\\nonumber
&&K_{10}=Ric\,\nabla^2\,Ric\quad,\quad K_{11}=Riem\,\nabla^2\,Riem\quad,\quad K_{12}=Ric\,\nabla\nabla R\quad,\quad K_{13}=|\nabla Ric|^2
\\\nonumber \\
&&K_{14}=\nabla Ric \nabla Ric\quad,\quad K_{15}=|\nabla Riem|^2\quad,\quad K_{16}=\nabla^2R^2\quad,\quad K_{17}=\nabla^4R~.
\end{eqnarray}
Explicit expansion of the trivial anomalies
\begin{eqnarray}
M_5&=& 6K_6-3K_7+ 12K_8+K_{10}-7K_{11}-11K_{13}+ 12K_{14}-4K_{15}~,\\
M_6&=& -\frac{1}{5} K_9  + K_{10} + \frac{2}{5} K_{12} + K_{13}~,\nonumber \\
M_7 &=& K_4 + K_5 - \frac{3}{20} K_9 + \frac{4}{5} K_{12} + K_{14}~,\nonumber\\
M_8 &=& -\frac{1}{5} K_9 + K_{11} + \frac{2}{5} K_{12} + K_{15}~,\nonumber\\
M_9 &=& K_{16} \nonumber\\
M_{10} &=& K_{17} \nonumber.
\end{eqnarray}
Corresponding conformal primitives~\footnote{These relations were reported in appendix C, eqn.41, of~\cite{Bastianelli:2000rs}. We are indebted to F. Bastianelli for pointing this out to us.}
\begin{eqnarray}
\nonumber
    \mathcal{F}_5=\frac{1}{30}K_1-\frac{1}{4}K_2+K_6\quad,\quad\mathcal{F}_6=\frac{1}{100}K_1-\frac{1}{20}K_2\quad,\\
\nonumber
    \mathcal{F}_7=\frac{37}{6000}K_1-\frac{7}{150}K_2+\frac{1}{75}K_3-\frac{1}{10}K_5-\frac{1}{15}K_6\quad,\quad\mathcal{F}_8=\frac{1}{150}K_1-\frac{1}{20}K_3\quad,\\
    \mathcal{F}_9=-\frac{1}{30}K_1\quad,\quad \mathcal{F}_{10}=\frac{1}{300}K_1-\frac{1}{20}K_9\quad.
\end{eqnarray}
Relevant curvature invariants at $S^1_{\beta}\times S^5$
\begin{eqnarray}
\nonumber
    K_1=R^3=8000\quad,\quad K_2=R\, Ric^2=1600\quad,\quad K_3=R\, Riem^2=800\quad,\\\nonumber\\
K_5=-Riem\,Ric^2=-320\quad,\quad K_6=Ric\,Riem^2=160\quad,\quad K_9=R\,\nabla^2\,R=0\quad.
\end{eqnarray}

\section{6D Branson's basis of conformal primitives}
On conformally flat manifolds the general basis of six trivial total derivatives is reduced to only three independent terms, stemming from only three conformal primitives. One convenient choice of conformal primitives consists of $J^3, |\nabla J|^2$ and $J|V|^2$.  
To compute the trivial total derivatives obtained by the conformal variation of these curvature invariants under $g\rightarrow g_{\omega}=e^{2\omega}g$,  it is enough to keep only the term linear in the conformal factor $\omega$ and discard boundary terms since the manifold is closed.\\
Let us start with $J^3$ 
\begin{equation}
    \int_{\hat{M}_6}dv_{\omega} J_{\omega}^3=\int_{M_6}dv (J-\nabla^2\omega)^3=\int_{M_6}dv (J^3-3J^2\nabla^2\omega)\;\rightarrow\;\int_{M_6}dv\, \omega (-3\nabla^2J^2)~,   
\end{equation}
so that $J^3$ turns out to be a conformal primitive for $-3\nabla^2J^2$.\\
In the case of $|\nabla J|^2$, as conformal primitive it is equivalent to $-J\nabla^2J$ since they add up to a boundary term $\frac{1}{2}\nabla^2J^2$. A shortcut to compute the conformal variation is provided by considering the Yamabe operator $Y=-\nabla^2 + 2J$ instead of the plain Laplacian, and exploiting the conformal covariance of $Y$ as follows
\begin{eqnarray}
    \nonumber
    \int_{\hat{M}_6}dv_{\omega} J_{\omega}Y_{\omega}J_{\omega}=\int_{M_6}dv (J-\nabla^2\omega)&Y&(J-\nabla^2\omega)=\int_{M_6}dv \left\{JYJ-(\nabla^2\omega)YJ-JY\nabla^2\omega)\right\}
    \\
    \;&\rightarrow&\;\int_{M_6}dv\, \omega (-2\nabla^2YJ)~. 
\end{eqnarray}
The integration by parts performed above is equivalent to the self-adjointness of both the Laplacian and the Yamabe operator. We end up with the conclusion that a conformal primitive for $-2\nabla^2YJ$ is given by $JYJ=J(-\nabla^2+2J)J$ or, equivalently, $|\nabla J|^2+2J^3$. Combining this with the previous result for $J^3$, we obtain that $|\nabla J|^2$ is a conformal primitive for $2\nabla^2\nabla^2J+2\nabla^2J^2$.\\
A shortcut to compute the conformal variation of the remaining curvature invariant $J|V|^2$ involves the Paneitz operator P acting on the Schouten scalar 
\begin{eqnarray}
    \int_{\hat{M}_6}dv_{\omega} Q_{4,\omega}J_{\omega}  =   \int_{\hat{M}_6}dv_{\omega} P_{\omega}J_{\omega} \nonumber &=& \int_{M_6}dv e^{\omega}Pe^{-\omega} (J-\nabla^2\omega) \nonumber \\ = \int_{M_6}dv \left\{PJ-P\nabla^2\omega+e^{\omega}P^o(e^{-\omega}J)\right\}&=& \int_{M_6}dv \left\{Q_4J-Q_4\nabla^2\omega+JP^o\omega\right\} \nonumber \\
    \;&\rightarrow&\;\int_{M_6}dv\, \omega (-\nabla^2Q_4+P^oJ)~. 
\end{eqnarray}
We have kept only terms at most linear in $\omega$ and made use of the following properties of the Q-curvature $Q_4$ and Paneitz operator $P$ in 6D
\begin{equation}
    P=P^{o}+Q_4\;,\; Q_4=3J^2-2|V|^2-\nabla^2J\;,\; P_{\omega}=e^{-5\omega}Pe^{\omega},  
\end{equation}
where $P^o$ denotes the total derivative part of the Paneitz operator 
\begin{equation}
    P^{o}=\nabla^2\nabla^2+4\nabla\cdot\left(V\cdot\;-\;J\right)\nabla~. 
\end{equation}
We have therefore that $Q_4J$ is a conformal primitive of $-\nabla^2Q_4+P^oJ$ which equals $2\nabla^2\nabla^2J-5\nabla^2J^2+2\nabla^2|V|^2+4\nabla\cdot(V\cdot\nabla J)$.
With all the instances gathered so far, we can finally disentangle the $-J|V|^2$ term in $Q_4J$, and obtain that it is a conformal primitive for $\nabla^2|V|^2+2\nabla\cdot(V\cdot\nabla J)+\nabla^2J$ .
We summarize the above results, which have also been independently verified with the {\it xAct} package for {\it Mathematica}, in the table below.
\begin{center}
\begin{tabular}{|c|c|}
\hline
    &  \\
    \qquad trivial total derivative \qquad  & \qquad conformal primitive \qquad\\& \\\hline
    & \\  
   $\nabla^2\nabla^2J$ & $\frac{1}{2}|\nabla J|^2 +\frac{1}{3}J^3$ \\&   \\\hline
    &   \\ $\nabla^2 J^2$ & $-\frac{1}{3}J^3$\\ &  \\\hline & \\  
   $\nabla^2|V|^2+2\nabla\cdot(V\cdot\nabla J)$ & $-J|V|^2+\frac{1}{3}J^3$\\&   \\\hline
\end{tabular}
\end{center}
Endowed with this correspondence, we can now determine the contribution to the partition function that originates in the trivial total derivatives of the trace anomaly on conformally flat backgrounds
\begin{eqnarray}
\nonumber (4\pi)^3\cdot\langle T \rangle &=& -a_6 \cdot E_6 + \gamma_1\nabla^2\nabla^2J + \gamma_2\nabla^2J^2+\gamma_3\{\nabla^2|V|^2+2\nabla\cdot(V\cdot\nabla J)\}\\\nonumber\\&=&-6\,a_6 \cdot Q_6 + \tilde{\gamma}_1\nabla^2\nabla^2J + \tilde{\gamma}_2\nabla^2J^2+\tilde{\gamma}_3\{\nabla^2|V|^2+2\nabla\cdot(V\cdot\nabla J)\}~.
\end{eqnarray}
The contribution to the partition function from the trivial total derivatives reduces to the evaluation on the cylinder of their corresponding conformal primitives. Collecting similar terms, evaluating the conformal primitives $J^3=8$, $|\nabla J|^2=0$ and $J|V|^2=3$, we arrive at volume times a particular combination of coefficients
\begin{equation}
    \frac{\beta\,\pi^3}{64\pi^3}\left\{ \frac{8\gamma_1}{3}-\frac{8\gamma_2}{3}+\frac{8\gamma_3}{3} -3\gamma_3\right\}=\frac{\beta}{24}\left\{ \gamma_1-\gamma_2-\frac{1}{8}\gamma_3 \right\}~,
\end{equation}
so that we obtain for the Casimir energy in the Euler basis 
\begin{equation}
    E_o=-\frac{15}{8}a_6-\frac{1}{24}\left(\gamma_1-\gamma_2 -\frac{1}{8}\gamma_3\right)~,
\end{equation}
or, alternatively, in the Q-curvature basis
\begin{equation}
    E_o=-3a_6-\frac{1}{24}\left(\tilde{\gamma}_1-\tilde{\gamma}_2 -\frac{1}{8}\tilde{\gamma}_3\right)~.
\end{equation}
The degeneracy of the basis for total derivatives at conformally flat manifolds is summarized in the table below. 
\begin{center}
\begin{tabular}{|c|c|}
\hline
    &  \\
    \qquad general basis of t.t.d. \qquad  & \qquad reduced basis of t.t.d. \qquad\\& \\\hline
    & \\  
   $M_5$ & $-24\{\nabla^2|V|^2+2\nabla\cdot(V\cdot\nabla J)\}-7\nabla^2J^2$ \\&   \\\hline
    & \\  
   $M_6$ & $8\{\nabla^2|V|^2+2\nabla\cdot(V\cdot\nabla J)\}-\nabla^2J^2$ \\&   \\\hline
     &   \\ $M_7$ & $8\{\nabla^2|V|^2+2\nabla\cdot(V\cdot\nabla J)\}-\frac{7}{2}\nabla^2J^2$\\ &  \\\hline & \\  
   $M_8$ & $8\{\nabla^2|V|^2+2\nabla\cdot(V\cdot\nabla J)\}-6\nabla^2J^2$\\&   \\\hline
      &   \\ $M_9$ & $100\nabla^2J^2$\\ &  \\\hline & \\  
   $M_{10}$ & $10\nabla^2\nabla^2J$\\&   \\\hline
\end{tabular}
\end{center}
This dictionary allows a direct verification of the vanishing on conformally flat manifolds of the freedom in Anselmi's pondered Euler density (eqn.\ref{Anselmi}) and the two freedoms found out by Hamada (eqns.\ref{Hamada1} and \ref{Hamada2}). It also allows to verify the relation for the Casimir energy in terms of the coefficients of the reduced basis, collecting their coefficients 
\begin{eqnarray}
    \nonumber E_o&=&-\frac{15}{8}a_6-\frac{1}{24}\left([10g_{10}]-[-7g_5-g_6-\frac{7}{2}g_7-6g_8+100g_9] -\frac{1}{8}[-24g_5+8g_6+8g_7+8g_8]\right)\\\nonumber\\
    &=&-\frac{15}{8}a_6-\frac{5}{12}\left(g_5+\frac{1}{4}g_7+\frac{1}{2}g_8-10g_9+g_{10}\right)~.
\end{eqnarray}

\section{Critical GJMS on the cylinder}
Let us show here that the evaluation of the dilaton action (eqn.\ref{Dilatonaction}) can be done as well on the cylinder. Again, the critical Q-curvature vanishes for topological reasons (vanishing Euler characteristic) whereas the critical GJMS gets factorized (see, e.g.~\cite{juhl2010conformally,Beccaria_2016,Beccaria:2017dmw,Aros:2023hgi})  
\begin{equation}
P_{n}=\prod_{j=1}^{n/2}\left\{-\partial_o^{\,2}\,+\,(\sqrt{\Delta_0}+2j-n/2-1)^2\right\}~,
\end{equation}
where the shifted Laplacian on the sphere $\Delta_0=-\nabla^2+\frac{(n-2)^2}{4}$ acting on the Weyl factor produces only a constant factor. Therefore the $\sigma\,\hat{P}_n\sigma$ term results simply in a factor $\prod_{j=2}^{n/2}\left\{2j-2\right\}$ times $-u\,\partial_o^2 u$. After integration by parts, we obtain again the contribution $((n-2)!!)^2$ times $\int du$ or $\beta$. We notice that the Weyl factor is a zero mode of the Laplacian, and only after integration by parts to get a kinetic term it does produce a nonvanishing contribution proportional the type A central charge $a$. The corresponding boundary term that is being discarded, is related to the fact that the mapping from flat space to the cylinder is an improper one. A more detailed analysis for a cylinder with finite height and careful consideration of boundary conditions should shed light on this subtlety. After all, the Polyakov formula in two dimensions relates a torus to another torus in the same conformal family, and not to a disk. Nonetheless, the finite contribution from the conformal primitives on the flat shell and the cylinder remains unaffected.

\section*{References}


\providecommand{\href}[2]{#2}\begingroup\raggedright\endgroup

\end{document}